\documentclass[aps,pra,reprint,groupedaddress]{revtex4-1}

\input{epsf}

\usepackage[pdftex]{graphicx}
\usepackage{amsmath} 

\begin{document}




\author{G.~G.~Kozlov}
\affiliation{Spin-Optics laboratory, St.~Petersburg State University, 198504 St.~Petersburg} 




\begin{abstract}
The signal registered by a plane photodetector placed behind an optically inhomogeneous object irradiated by two coherent Gaussian beams intersecting inside the object at small angle to each other is calculated in the single-scattering approximation.  In the considered arrangement, only one of the beams hits the detector and serves as local oscillator for heterodyning the field scattered by the other beam (not hitting the detector). The results of analytical calculation show that the signal detected in this way is contributed only by the region of the inhomogeneous object where the two beams overlap. By moving the scatterer with respect to the overlap region and monitoring the heterodyned signal, with the aid of the derived expression, one can reconstruct the refractive-index relief of the scatterer.  We also propose a simple method of spatial mapping of the sample that allows one to estimate the magnitude and characteristic dimensions of the inhomogeneities.   
 \end{abstract}

\title{Heterodyne detection of scattered light: Application to mapping and tomography of optically inhomogeneous media}
\maketitle

\section{Introduction  }

Heterodyning is known to be an efficient method of detecting weak signals. In a simplest case, the heterodyning implies summation of a weak signal $E_1$ to be detected with a strong signal of fixed amplitude $E_0$ created by the {\it local oscillator} with subsequent measurement of the of the obtained sum ($S$) squared: $S=E_0^2+2E_0E_1+E_1^2$. Under these conditions, the contribution linear in $E_1$ (carrying all the information about the weak signal $E_1$) proves to be proportional to the controllable amplitude $E_0$ that may be increased, thus increasing sensitivity of detecting the signal $E_1$. The heterodyning method is widely used nowadays in radio-electronics, microwave technique, and optics \cite{Bor,For,Step,Adam,Atlan}.  In optics, for the heterodyne detection of a weak field $E_1$ using conventional photodetectors (photodiodes or photomulotipliers), it suffices to apply a strong field $E_0$ to the same detector. Then, the output signal of the detector $S$ proportional to total intensity $I$ of the detected field will contain the above contribution bilinear in the field amplitudes $S\sim I\sim (E_1+E_0)^2=E_0^2+2E_0E_1+...$.

One important problem that is often solved with the aid of heterodyning is related to tomography, which implies detection of optical fields arising upon scattering of laser beams in an inhomogeneous medium with subsequent restoration of spatial relief of the inhomogeneity \cite{T1,T2,Huang,Lee,Enomoto}. Specific feature of optical heterodyning is that  dimensions  of photosensitive surface of the photodetector, as a rule, considerably exceed the light wavelength, and, therefore, when calculating the output signal of the detector, one has to take into account the effects of spatial interference of the fields of signal and local oscillator.  

 In this work, we present analysis of heterodyne detection of optical scattering in the two-beam arrangement \cite{Sht,Rud,Sht1} of collinear heterodyning \cite{Carl}.  In this arrangement, the two beams (the main and the tilted), intersecting at a small angle $\Theta< 1$ rad at some point inside the sample, are obtained from the same laser.  The photodetector, in its chosen position, directly detects only the main beam transmitted through the sample. Under these conditions, this beam plays the role of the local oscillator needed to detect the scattered light that also hits the photodetector (Fig. \ref{fig1}).  In papers \cite{Sht,Rud,Sht1}, the two-beam arrangement was used for 3D recording and reading of information. In those studies, for spatial selection of the recorded holograms, frequencies of the beams were different.  
 
 In the present paper, we analyze possibility of using such an arrangement for tomography and mapping of scattering media.  We show that it is possible to use, for these purpose, the beams of the same frequency obtained from the same laser source (what is often called {\it homodyning}). The analysis is based on Eq. ((\ref{18})) (see below) that associates the heterodyned scattering signal with spatial overlap of the two beams.  In spite of the fact that the collinear heterodyning is well known and has been actively studied earlier, we did not manage to find in the literature the simple equation ((\ref{18})) that allows one to formulate and solve the problem of tomography of optically inhomogeneous transparent media. 
 
 It is interesting to note similarity between such experiments on light scattering and those of the spin-moise spectroscopy – a new direction of research developed during the last decade \cite{Alex,Croo,Oest1,Oest2,zap1}. The signal formation in the spin noise spectroscopy can be considered as heterodyning of the field scattered on fluctuations of gyrotropy  \cite{Gorb}.  The two-beam arrangement of the spin-noise experiments and its informative capabilities are described in \cite{Koz}.

The paper is organized as follows. In the theoretical part, we calculate, in the single-scattering approximation, the heterodyned scattering signal (HSS) detected in the above two-beam arrangement for the sample with weak inhomogeneity of its refractive index. We show [Eq. (\ref{18})] that, in the case of a linear weakly inhomogeneous medium, the HSS is formed only by the region of spatial overlap of the beams. For this reason, by moving the sample and detecting the HSS, it is possible to restore profile of susceptibility (refractive index) in the sample.  The calculations are illustrated in the experimental part of the paper. A simple setup for detection of the HSS described in this section makes it possible to perform mapping of inhomogeneous transparent objects.  We show that from the 2D images thus obtained one can evaluate characteristic dimensions and magnitude of optical inhomogeneities of the sample under study.

\section{Two-beam arrangement for detection of scattering: basic equations }

     \begin{figure}
       \includegraphics[width=.8\columnwidth,clip]{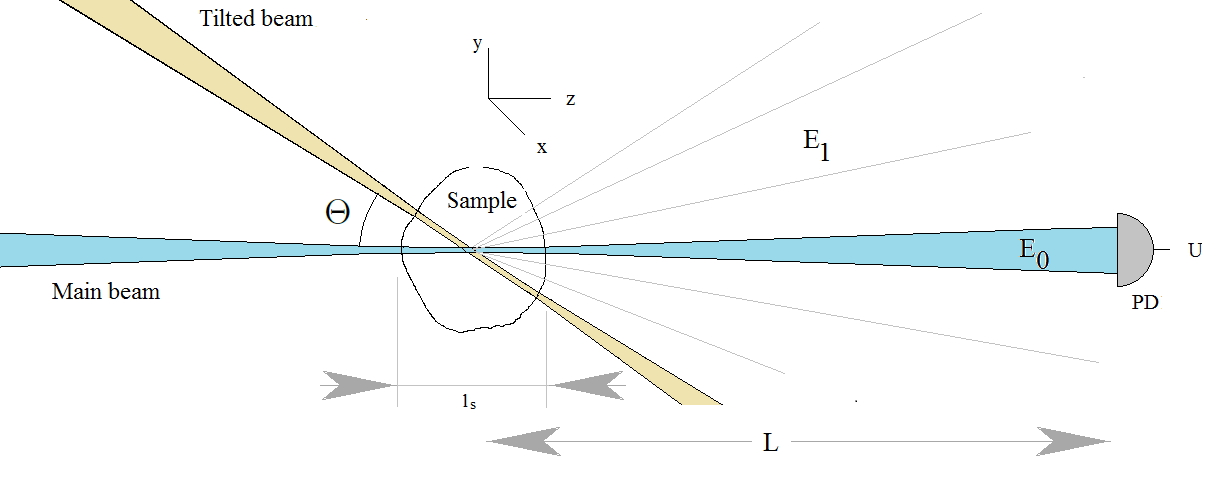}
        \caption{Two-beam arrangement for heterodyne detection of optical inhomogeneity}
        \label{fig1} 
       \end{figure}

 Schematic of the two-beam detection of the heteridyned signal scattering  (HSS) is shown in Fig. \ref{fig1}. The two coherent laser beams (the main and the tilted) with the frequency $\omega$ are incident on the sample and excite the scattered field ${\bf E_1(r)}$ which is registered by the photodetector PD located at a distance $L$ from the sample. The main beam, after passing through the sample, is also incident on the photodetector PD. The field ${\bf E_0(r)}$  of this beam plays the role of a local oscillator for detection of the scattered field ${\bf E_1(r)}$.
 
 In what follows, we will use the complex electromagnetic fields dependent on time as $ e^{-\imath\omega t}$, assigning physical sense only to their real parts, which will be denoted by the corresponding calligraphic characters. In addition, we will assume, in the calculations, that the scattering sample is positioned near the coordinate origin and has a characteristic size $l_s$ much smaller than the distance to the photodetector $L$.

In the chosen coordinate system, the $xy$ plane is aligned parallel to the photosensitive surface of the photodetector PD, and the axis $z$ is collinear with the main beam propagation direction.  Let us define the output signal $U$ of the photodetector PD as a square of the total electric field ${\bf E(r)=E_1(r)+E_0(r)}$ on the surface of the photodetector averaged over the period $2\pi/\omega$ of the optical oscillation    and integrated over the photosensitive surface of the detector $S$:

 \begin{equation}
   \begin{split}
   U={\omega\over 2\pi}\int _0^{2\pi/\omega}dt\int_S dxdy \hskip1mm [\hbox{Re }E(x,y,L)]^2\equiv
\\ \equiv {\omega\over 2\pi}\int _0^{2\pi/\omega}dt\int_S dxdy \hskip1mm [{\cal E}(x,y,L)]^2
\end{split}
\label{1}   
\end{equation}
The HSS we are interested in is the contribution $\delta U$ into the output signal of the photodetector linear in the scattered field strength. One can see that this contribution is given by the expression
   \begin{equation}
   \begin{split}
     \delta U={\omega\over \pi}\int _0^{2\pi/\omega}dt\int_S dxdy
     \bigg [
        {\cal E}_{1x}(x,y,L){\cal E}_{0x}(x,y,L)+
        \\ +{\cal E}_{1y}(x,y,L){\cal E}_{0y}(x,y,L)
        \bigg ]
        \end{split}
        \label{2}
        \end{equation}
         
   Here, in conformity with the notations accepted above, ${\cal E}_0(x,y,L)\equiv $ Re ${\bf E_0}(x,y,L)$ and  ${\cal E}_1(x,y,L)\equiv $ Re ${\bf E_1}(x,y,L)$. Thus, to calculate the HSS, we have to find the field $\bf E_1(r)$, using the fields of the main ${\bf E_0( r})$  and the tilted ${\bf E_0^t(\bf r})$ beams, to take its real part ${\cal E}_1({\bf r})$, and to calculate integrals (\ref{2}).
   
   Now, we present expressions for electric fields of the main and tilted beams that we will further use in our calculations.  Assuming that the main beam is Gaussian and propagates along the $z$-axis, we can use the expression for the electric field of such a beam from \cite{Koz}:
   \begin{widetext}
    \begin{equation}
                     {\bf E_0(r)}=e^{\imath (kz-\omega t)}  
                     \sqrt{8 W\over c}
                     {kQ\over (2k+\imath Q^2z)}
                     \exp\bigg [-{kQ^2(x^2+y^2)\over 2(2k+\imath Q^2z)}\bigg ]
                     {\bf d}\equiv {\bf A_0(r)} e^{-\imath\omega t},
                     \label{3}
                     \end{equation}
    \end{widetext}
    where ${\bf r}=(x,y,z)$. Polarization of the beam is specified by the unit Jones vector ${\bf d}$,  and, for the beam (\ref{3}), this vector has only $x$ and $y$ components \cite{note1}.  For definiteness, we assume that the main beam is polarized linearly along the $x$:  ${\bf d}=(1,0,0)$.  The field of the tilted beam, ${\bf E_0^t(r)}$, can be obtained from that of the main beam by rotation around the $x$ - axis by a small angle $\Theta$ with a shift $\delta {\bf r}\equiv (\delta x, \delta y, \delta z)$:
                     
        \begin{widetext}                   
                 \begin{equation}
                  {\bf E_0^t(r)}=e^{\imath (kZ-\omega t+\phi_t)}  
                  \sqrt{8 W_t\over c}
                  {kQ\over (2k+\imath Q^2Z)}
                  \exp\bigg [-{kQ^2(X^2+Y^2)\over 2(2k+\imath Q^2Z)}\bigg ]
                  {\bf d^t} \equiv {\bf A_0 ^t(r)} e^{-\imath\omega t}
                  \label{4}
                  \end{equation}
                  \end{widetext}
where
                              \begin{equation}
                              \left(\begin{matrix} X\cr Y\cr Z \end{matrix}\right )\equiv\hat R {\bf r}+{\bf \delta r},  \hskip5mm 
                              \hat R\equiv \left(\begin{matrix}1 & 0& 0 \cr 0 & \cos\Theta &\sin\Theta \cr 0 & -\sin\Theta & \cos\Theta \end{matrix}\right )
                              \label{5}
                              \end{equation}

The phase of the tilted beam may be shifted with respect to that of the main beam (e.g., using the time delay line). This shift is denoted by $\phi_t$.  Polarization of the tilted beam is specified by the unit vector  ${\bf d^t}$ that can be chosen with no regard for ${\bf d}$ (retaining transversality of the beam field). At small $\Theta$, the tilted beam evidently also has only $x$- and $y$- components. In Eqs. (\ref{3}) and (\ref{4}),  the quantities $W$ and $W_t$ are the intensities of the main and tilted beams, $c$ is the speed of light, and $Q\equiv 2/\rho_c$, where $\rho_c$ is the beam radius on the e-level of the field in the beam waist.  Besides, in Eqs. (\ref{3}) and (\ref{4}), we introduced time-independent amplitudes of the fields ${\bf A_0(r)}$ and ${\bf A_0^t(r)}$.

\section{Calculating the HSS in the single-scattering approximation}

  Let us pass to mathematical formulation of the scattering problem that should be solved to employ Eq. (\ref{2}). The scattering sample is supposed to be characterized by a spatially inhomogeneous polarizability $\alpha ({\bf r})$, which will be considered scalar \cite{note2}  and small $|\alpha ({\bf r})|\ll 1$.  This function is nonzero in the spatial region whose characteristic dimensions $l_s$ are considered to be small as compared with  the distance $L$ from the sample to the detector, $L\gg l_s$ (Fig.\ref{fig1}).  In this case, when the electromagnetic field varies with time as $\sim e^{-\imath\omega t}$, Maxwell’s equations lead to the following expressions for the electric field and polarization     
     
\begin{equation}
\begin{split}
\Delta {\bf E}+k^2{\bf E}=-4\pi k^2{\bf P}-4\pi \hbox{ grad div }{\bf P}, \\ 
{\bf P(r)}=\alpha({\bf r}){\bf E(r)}
\end{split}
\label{6}
\end{equation}

 where $k={\omega/c}={2\pi/\lambda}$ ($\lambda $ is the wavelength of the light with the frequency $\omega $). In the single-scattering approximation, solution of this equation is usually represented in the form of power series over $\alpha ({\bf r})$, retaining only terms of zeroth and first order (which is possible when $|\alpha ({\bf r})|\ll 1$).  As the zero-order terms, one should take the fields of the main and tilted beams. Then, for the part of the scattered field ${\bf E_1(r)}$ arising due to scattering of the tilted beam \cite{note3}, one can easily obtain the following inhomogeneous Helmholеz equation $\Delta {\bf E_1}+k^2{\bf E_1}=-4\pi k^2\alpha({\bf r}){\bf E_0^t(r)}-4\pi \hbox{ grad div }\alpha({\bf r}){\bf E_0^t(r)}$,  with its right-hand side representing a sum of two terms.  We will perform calculations only for the first term that, at small angles $\Theta $, provides the main contribution to the HSS. The role of the second term, in this case, proves to be small, which can be ascertained by making calculations similar to those presented below.  As will be shown below, the HSS is controlled only by the region of the sample where the main and tilted beams overlap.  In our experiments, this is the region of overlap of the main and tilted beam waists.  In our experiments, for the typical angles $\Theta\sim 0.1 - 0.2$ rad, dimensions of this region did not exceed the Rayleigh length $z_c\equiv \pi\rho_c^2/\lambda=4\pi /Q^2\lambda$, which was about 2 mm.  For this reason, for the experiments described below, we may assume that the HSS is formed by the region of overlap of the beam waists in the sample {Footnote 4: Regions of quasi-cylindrical shape}.  Thus, to find the scattered field ${\bf E_1(r)}$ produced by the tilted beam we have to solve the inhomogeneous Helmholtz equation

 \begin{equation}
 \Delta {\bf E_1}+k^2{\bf E_1}=-4\pi k^2\alpha({\bf r}){\bf E_0^t(r)}\equiv -4\pi k^2 {\bf P^t(r)} .
 \label{7}
 \end{equation}
  Solution of this equation can be obtained using Green’s function $ \Gamma({\bf r}) $ of the Helmholtz operator: $ \Gamma({\bf r})=-{e^{\imath kr}/4\pi r}$      and has the following form
     \begin{equation}
      {\bf E_1(r)}=k^2\int  {e^ {\imath k|{\bf r-R}|}\over {\bf |r-R|}}  {\bf P^t(R)}
      d^3{\bf R}
      \label{8}
       \end{equation}
  For further calculations, it is convenient to introduce the function $\Phi({\bf R})$ defined by the equation:
 \begin{equation}
 \Phi({\bf R})\equiv \int_S dxdy\hskip2mm 
         {\cal E}_{0x}(x,y,z)
          {e^ {\imath k|{\bf r-R}|}\over {\bf |r-R|}}\bigg |_{z=L} 
 \label{9}
 \end{equation}

  Here, the integration is performed over the surface of the photodetector, whose dimensions we assume to be large as compared with the size of the main beam spot on the detector. This allows us, in the calculations, to consider the integration limits infinite. The function $\Phi({\bf R})$  has the sense of the field created by a flat polarized layer located on the surface of the detector $S$, with the spatial dependence of “polarization” of this layer being controlled by the field of the main beam  ${\cal E}_{0x}(x,y,L)$ on he surface of the detector. This is why we can suppose that the field $\Phi({\bf R})$ will be similar to that of the main beam and hence, will represent the beam converging at the coordinate origin   \cite{note4}. 
  Using Eq. (\ref{2}), one can easily show that the observed HSS $\delta U$ is expressed through the introduced function $\Phi ({\bf R})$ as follows
  
  \begin{equation}
  \delta U=k^2\hbox { Re }{\omega\over \pi}\int _0^{2\pi/\omega}dt\int d^3{\bf R}\hskip2mm\Phi({\bf R}) 
          P^t_x({\bf R})
          \label{10}
  \end{equation} 
 
  It follows from this equation that the HSS $\delta U$ is determined by overlap of the field $\Phi({\bf R})$  (which, as we suppose, is similar to the field of the main beam ${\cal E}_{0x}({\bf R})$) with the field of the tilted beam ${\bf E_0^t(R)}$  (since $P^t_x({\bf R})=\alpha ({\bf R})E^t_{0x}({\bf R})$)). Let us calculate the function $\Phi({\bf R})$ (\ref {9}) at large $L$.  When the size of the main beam spot on the photodetector surface (this size can be calculated as $LQ/k=L\lambda/\pi\rho_c$) is much smaller than $L$, the real part of the field of the main beam on this surface ${\cal E}_{0x}(x,y,L)$ is given by the expression
 \begin{widetext}
 \begin{equation}
                  {{\cal E}_{x0}(x,y,L)}=\sqrt{8 W\over c}
                {k\over  QL}\hskip1mm \sin \bigg [ kL-\omega t
                +{k[x^2+y^2]\over 2L}
                \bigg ]  
                                                  \exp\bigg [-{k^2(x^2+y^2)\over  Q^2L^2}\bigg ],
                                                  \label{11}
                                                  \end{equation}
 \end{widetext}
 which can be obtained from Eq. (\ref{3}) at $|x|,|y|\ll L$. Using the fact that $L$ is much larger than all dimensions of the problem, we can simplify Eq. (\ref{9}) for $\Phi({\bf R})$ and distinguish explicitly the factors $e^{\pm\imath\omega t}$:
  \begin{equation}
  \begin{split}
 \Phi({\bf R})= {1\over L}\int_S dxdy\hskip2mm 
         {\cal E}_{0x}(x,y,z)
          e^ {\imath k|{\bf r-R}|}\bigg |_{z=L} \equiv \\
          \equiv e^{-\imath\omega t}\Phi_+({\bf R})+e^{\imath\omega t}\Phi_-({\bf R})
 \end{split}
 \label{12}
 \end{equation}

Below we will need only the function $\Phi_-({\bf R})$ since $P^t_x({\bf R})\sim e^{-\imath\omega t}$, and only $\Phi_-({\bf R})$ will survive upon averaging over the light wave period in Eq. (\ref{10}).
Using Eqs. (\ref{9}) and (\ref{12}), we can obtain, for the function $\Phi_-({\bf R})$, the expression

\begin{equation}
\Phi_-({\bf R})=-\sqrt{8 W\over c}{k\over  2\imath QL^2}e^{-\imath kL}\hskip1mm I_-
\end{equation}
 where $I_-$  represents the following integral  
\begin{widetext}
 \begin{equation}
  I_-=L^2\int d\xi d\eta \exp N\bigg [ \imath \sqrt{
  \xi^2+\eta^2+(1-R_z/L)^2
  }-w([\xi+\rho_x]^2+[\eta+\rho_y] ^2)\bigg ]
  \label{14}
  \end{equation}
\end{widetext}
  
  Here, we introduced the following notations $\xi \equiv x/L, \eta\equiv y/L, \rho_x\equiv R_x/L,\rho_y\equiv R_y/L, N\equiv kL\gg 1, w\equiv \imath/2 +k/LQ^2=\imath/2+z_c/2L$ (where $z_c=\pi\rho_c^2/\lambda$ -- is the Rayleigh length, $\rho_c$ -- is the beam radius on e-level of its waist). It can be shown that $I_-$ depends on $\rho\equiv \sqrt{\rho_x^2+\rho_y^2}$, and, upon integration over the sample volume in Eq. (\ref{10}), the following condition is satisfied $\rho\sim l_s/L\ll 1$. The integrand in Eq. (\ref{14}) is essentially nonzero in the region with dimensions of about $|N \hbox { Re } w|^{-1/2}=\lambda/\pi\rho_c\ll 1$.  Since $R_z/L\sim l_s/L\ll 1$, the quantity $H\equiv 1-R_z/L$ is $\sim 1 $. These estimates show that the square root in Eq. (\ref{14}) can be expanded as follows: $\sqrt{\xi^2+\eta^2+ {(1-R_z/L)^2}}=H+[\xi^2+\eta^2]/ 2H$.  After that, the integral (\ref{14}) is reduced  to the product of two independent Gaussian integrals in $\xi$ and $\eta$, with each of them calculated using the formula
  \begin{equation}
               \int dx \exp [-\alpha x^2+\beta x ]=\sqrt{\pi\over \alpha}\exp \bigg ({\beta^2\over 4\alpha}\bigg ).
                      \label{15}
              \end{equation} 
   This formula is valid at arbitrary complex $\beta$ and at Re $\alpha >0$. We see that, as was supposed above, the function $\Phi_-({\bf R})$ is expressed through the main beam amplitude $A_{0x}(\bf R)$:
    
   \begin{equation}
    \Phi_-({\bf R})=
    -{\imath \pi\over k}\hskip1mm A_{0x}^\ast({\bf R})\hskip10mm |{\bf R}|\ll L
    \label{16}
    \end{equation}
By substituting this to Eq. (\ref{10}), we come to the following expression for the HSS:
   \begin{equation}
     \delta U= 2\pi k\hbox { Im }\int d^3{\bf r}\hskip2mm A_{0x}^\ast ({\bf r}) 
                              \alpha({\bf r}) A_{0x}^t{\bf(r)}  
    \end{equation} 
     where $A_{0x}({\bf r})$ is given by Eq. (\ref{3}).  In the above calculation of the HSS, the susceptibility $\alpha ({\bf r})$ was assumed scalar and the main beam polarized along the $x$-axis.  In the general case of tensor susceptibility and arbitrarily polarized main beam, similar calculations give the following general expression for the HSS:
   
      \begin{equation}
           \delta U=   2\pi k\hbox { Im }\int d^3{\bf r}\hskip2mm ({\bf A_{0} ( r}), 
                                    \hat \alpha({\bf r}) {\bf A_0^t(r)}),  
          \label{18}
          \end{equation}

    where the amplitudes ${\bf A_0(r)}$ and ${\bf A_0^t(r)}$  can be calculated using Eqs. (\ref{3}) and (\ref{4}),  with the scalar product given by the standard relation $({\bf A_{0} }, \hat \alpha  {\bf A_0^t})\equiv A_{0i}^\ast\alpha_{ik}A_{0k}^t$. In the considered case of small angles $\Theta$ between the main and tilted beams, in the scalar product of Eq. (\ref{18}), we can leave only $x$- and $y$-components of the vectors playing, in this case, the main role.

In the above treatment, the tilted beam (creating the scattered field) and the main beam (playing the role of local oscillator upon detection of the HSS) were considered to be independent \cite{note5}.  Let us show now that, in the case of transparent scatterer, the HSS created by the main beam proper vanishes.  To calculate this HSS, we have to set the tilted beam field equal to that of the main beam ${\bf E_0(r)\equiv E^t_0(r)}$.  For a transparent scatterer, the  polarizability tensor is Hermitian $\alpha_{ik}({\bf r})=\alpha ^\ast_{ki}({\bf r})$, and, therefore, the quadratic form $({\bf A_{0} ( r}), \hat \alpha({\bf r}) {\bf A_0 (r)})$ entering Eq. (\ref{18}) is always real, and $\delta U=0$.  This result could be anticipated, because the body inserted into the beam (even transparent) can only diminish intensity of the optical field on the detector placed behind this body.  For this reason, this intensity should be {\it even} function of $\alpha({\bf r})$ and, hence, in the single-scattering approximation (linear in $\alpha ({\bf r})$), the HSS should vanish.  In the next section, we will show that, by moving the sample with respect to the fixed (main and tilted) beams and detecting the HSS, it is possible (at least, in principle) to restore relief of susceptibility of the sample $\hat\alpha ({\bf r})$.

\section{Application of the HSS to tomography of transparent nongyrotropic objects. }

Consider possibility of application of the above approach for optical tomography, i.e., for restoration of spatial relief of optical susceptibility of inhomogeneous samples (scatterers). The calculations will be performed for the typical (in our experiments) values of $\rho_c=30 \mu$m and $\Theta\sim 0.1 - 0.2$ rad. In addition, it will be convenient to deal with the normalized HSS $\delta u\equiv \delta U/U_0$, where $U_0$ is the signal from the detector irradiated by the main beam. The signal $U_0$ is calculated using Eq. (\ref{1}) where 
${\cal E} (x,y,L) \rightarrow {\cal E}_{x0}(x,y,L)$, with  ${\cal E}_{x0}(x,y,L)$ defined by Eq. (\ref{11}).  Taking into account that aperture of the detector substantially exceeds the size of the main beam spot and integrating over $dxdy$,  within infinite limits, we obtain that $ U_0=2\pi W/ c$. 

We will restrict our treatment to the case of a transparent nongyrotropic scatterer, with the tensor $\hat\alpha ({\bf R})$ being symmetric and real Im  $\hat\alpha({\bf R})=0, \alpha_{ik}({\bf R})=\alpha_{ki}({\bf R})$. Let us fix positions of the main and tilted beams and replace the scatterer by the vector $-{\bf r}$. Then, the relief of the scatterer susceptibility will also be displaced $\hat\alpha ({\bf R})\rightarrow \hat\alpha ({\bf R+r})$ and, hence, the HSS will become a function of ${\bf r}$:
$\delta u\rightarrow \delta u({\bf r})$.  This function can be measured and used to restore the unknown function $\hat\alpha({\bf r})$ in the following way.  Let us introduce a real tensor  $\hat T({\bf R})$ defined by the equation

 \begin{equation}
T_{ki}({\bf R})\equiv { kc\over W} \hbox { Im }A_{0i}^\ast({\bf R}) A_{0k}^t({\bf R})
\end{equation} 
One can see that this tensor is essentially nonzero when ${\bf R}$  belongs to the region of overlap between the main and tilted beams. Using Eq. (\ref{18}), we can easily obtain for the HSS $\delta u({\bf r})$ the expression

\begin{equation}
\delta u({\bf r})=\int d{\bf r'} \hbox{ Sp }\hat\alpha({\bf r'})\hat T({\bf r'-r}),
\label{20}
\end{equation} 

which represents a convolution-type integral equation for the tensor function $\hat\alpha ({\bf r})$. By passing to the Fourier transform, this equation can be reduced to the algebraic one and, in certain cases, can be solved. 
 Consider, for example, the case, when the main and tilted beams are polarized along the $x$-axis. In this case, the tensor $\hat T({\bf r})$  has the only nonzero component $T_{xx}({\bf r})={ kc/2W} \hbox { Im } A_{0x}^\ast ({\bf r})A_{0x}^t({\bf r})$, and Eq. (\ref{20}) acquires the form

\begin{equation}
\delta u({\bf r})=\int d{\bf r'} \alpha_{xx}({\bf r'}) T_{xx}({\bf r'-r}),
\label{21}
\end{equation}  

We will denote the Fourier transforms of the functions entering this equation by letters with tilde.  For instance, $ T_{xx}({\bf r})=(2\pi)^{-3}\int e^{\imath {\bf qr}}\hskip1mm \tilde T_{xx}({\bf q})\hskip1mm d^3{\bf q}$.  Then, by passing in Eq. (\ref{21}) to Fourier transforms, we have 
\begin{equation}
\tilde\alpha_{xx}({\bf q})={\delta\tilde u({\bf q})\over \tilde T_{xx}(-{\bf q})}
\label{91}
\end{equation}
 
The region of overlap of the beams, under conditions typical for our experiments ($\rho_c=30 \mu$m and $\Theta \sim 0.1- 0.2$ rad.), appears to be essentially smaller that the Rayleigh length of the beams. For this reason, when calculating the component $T_{xx}({\bf r})$ of the tensor $\hat T({\bf r})$, the beams, in the region of their overlap, may be considered  as quasi-cylindrical, and we can write
\begin{widetext}
\begin{equation}
T_{xx}({\bf r}) = -{ 2kQ^2}\sqrt{W_t\over W}\hskip1mm \sin \bigg [kz {\Theta^2\over 2}+ky\Theta -\phi_t\bigg ]
 \exp \bigg [-{Q^2\over 2}\bigg ( x^2+y^2   +{z^2\Theta^2\over 2} +yz\Theta \bigg )\bigg ]
\label{24}
\end{equation}
\end{widetext}
We used here Eq. (\ref{5})  for $X,Y,Z$ at $\delta {\bf r}=0$  and small  $\Theta$. From Eq. (\ref{24}), one can see that the function $T_{xx}({\bf r})$ is essentially nonzero in the region with dimensions estimated as $4/Q\Theta=2\rho_c/\Theta$ (along the $z$-axis) and as $2\sqrt 2/Q=\sqrt 2\rho_c$ (in the $xy$-plane). For the beams used in our experiments, these quantities are, respectively, $\sim 300\mu$m $\times$ $ 50 \mu$m. Remind that the angle $\Theta$ should be sufficiently small for applicability of the employed approximations of trigonometric functions and sufficiently large to make sure that the region of beam ovelap does not exceed the Rayleigh length. This imposes the following conditions upon the angle: $1>\Theta>\lambda/\pi\rho_c$, which is well satisfied in our experiments. By performing the Fourier transform  of the function $T_{xx}({\bf r})$ at $\phi_t=0$, we obtain: 
\begin{widetext}  
  \begin{equation}
\tilde T_{xx}({\bf q})= {2\imath k (2\pi)^{3/2}\over Q\Theta}\sqrt{W_t\over W}\exp\bigg (
-{1\over Q^2}\bigg [{q_x^2\over 2}+\bigg ({2q_z\over \Theta}-q_y\bigg )^2\bigg ]
\bigg )
\bigg [\exp\bigg (-{[k\Theta+q_y ]^2\over 2Q^2}\bigg )- \exp\bigg (-{[k\Theta-q_y ]^2\over 2Q^2}\bigg )\bigg ]
\label{94}
\end{equation}  
\end{widetext}
 
 By measuring the spatial dependence of the HSS $\delta u({\bf r})$ and calculating its Fourier image $\delta\tilde u({\bf q})$, we can  calculate, using Eqs. (\ref{91}) и (\ref{94}), the quantity $\tilde\alpha_{xx} ({\bf q})$ and thus find spatial relief of $xx$-component of the susceptibility tensor of the sample: $\alpha_{xx}({\bf r})=(2\pi)^{-3}\int d^3{\bf q} e^{\imath {\bf qr}}\hskip1mm \tilde \alpha_{xx}({\bf q})$. 
 By rotating the polarization directions of the main and tilted beams by 90 degrees and making similar measurements of the HSS $\delta u({\bf r})$, we can obtain spatial relief of the $yy$-component of the susceptibility tensor $\alpha_{yy}({\bf r})$, and, finally, by measuring the HSS $\delta u({\bf r})$ with the main beam polarized along the $x$-axis and the tilted beam along the $y$-axis, we can restore spatial relief of the $xy$-component $\alpha_{xy}({\bf r})$. 
 
 In real experiments, we often have to distinguish between the HSS and different spurious signals. For that purpose, it is possible, e.g., to modulate intensity of the tilted beam (which does not hit the detector) and to  detect synchronous modulation of the detected signal. Our experience shows that it is most conveniently, for this purpose, to slightly modulate phase of the tilted beam $\phi _t=\phi_{t0}\sin\Omega t, \phi_{t0}<1$. This  can be made either with the aid of a mechanically variable delay or using a Pockels cell in the channel of the tilted beam. In this case, the observed modulation of the HSS (denote it $S({\bf r})$)) varies with time as 
 
\begin{equation}
 S({\bf r})\sin\Omega t= {\partial \delta u({\bf r})\over \partial \phi_t}   
 \hskip1mm \phi_{t0}\sin\Omega t
\end{equation}

and can be easily distinguished as a component of the output signal at the frequency  $\Omega$. When restoring the susceptibility relief with the aid of the $S({\bf r})$ signal, one has to use the tensor $\hat D\equiv \partial \hat T/\partial \phi _{t}$  and its Fourier transform
\begin{widetext}
 \begin{equation}
\tilde D_{xx}({\bf q})= 2{\phi_{t0} k}{(2\pi)^{3/2}\over Q\Theta}\sqrt{W_t\over W}\exp\bigg (
-{1\over Q^2}\bigg [{q_x^2\over 2}+\bigg ({2q_z\over \Theta}-q_y\bigg )^2\bigg ]
\bigg )
\bigg [\exp\bigg (-{[k\Theta+q_y ]^2\over 2Q^2}\bigg )+ \exp\bigg (-{[k\Theta-q_y ]^2\over 2Q^2}\bigg )\bigg ]
\label{97}
\end{equation}  
\end{widetext}

\section{ Mapping of thin samples}

When the sample is a plate normal to the $z$-axis, with its thickness $h$ so small that the changes of the functions $T_{xx}({\bf r})$ and $\hat\alpha ({\bf r})$ along the $z$-direction within the plate thickness can be neglected $T_{xx}(x,y,z)|_{z\in [-h/2,h/2]}\approx T_{xx}(x,y,0) $,   
 $\hat\alpha ({\bf r})\rightarrow \hat\alpha(x,y)$, then the above treatment can be simplified.  Let us introduce the susceptibility $a_{xx}(x,y)$ averaged over the beam along the $x$-direction

\begin{equation}
a_{xx}(x,y')\equiv {Q\over \sqrt{2\pi}}\int  \alpha_{xx}(x',y')e^{-Q^2(x-x')^2/2} \hskip1mm dx'
\label{28}
\end{equation}  
Then Eq. (\ref{21}) for the HSS obtained upon displacement of the plate with respect to the region of beam overlap along the $y$-axie can be rewritten as follows 
\begin{widetext}
 \begin{equation}
\delta u(x,y)= -2{h kQ}\sqrt{2\pi W_t\over W}\hskip1mm \int dy' a_{xx}(x,y') 
 \sin \bigg [k(y-y')\Theta -\phi_t\bigg ] \exp \bigg [-{Q^2(y-y')^2\over 2}     \bigg ] 
 \label{29}
\end{equation}
\end{widetext}

This is a one-dimensional convolution-type equation with respect to $a_{xx}(x,y)$, that, at a given $x$,   can be solved by passing to the Fourier transform over  $y$ variable.  The signal (\ref{29}) can be easily observed by placing the studied quasi-plane scatterer onto a vibrator oscillating along the $y$-axis.  By smoothly moving the vibrator with the scatterer along the $x$-axis, one can record the function $\delta u(x,y)$ and thus restore the susceptibility relief $\hat\alpha (x,y)$ as it was described above. 

In certain cases, however, one can get some idea about the character and magnitude of the inhomogeneity using direct mapping of the function $\delta u (x,y)$. Let us illustrate it by two model examples.  

As the first example, let us consider a small scatterer with the susceptibility 
$\alpha_0$ and area $S_0<(\lambda /\Theta)^2$, with its center at $x_0,y_0$.  In this case, the spatial dependence of the susceptibility can be presented in the form 
$\alpha_{xx}(x,y)=S_0\alpha_0\delta (x-x_0)\delta (y-y_0)$. Then, from Eqs. (\ref{28}) and   (\ref{29}), we can obtain, for the HSS detected from such a scatterer, the following expression:

 \begin{widetext}
 
   \begin{equation}
  \delta u(x,y)=- {2h kQ^2 \alpha_0 S_0}\sqrt{ W_t\over W}\hskip1mm  
   \sin \bigg [k(y-y_0)\Theta -\phi_t\bigg ] \exp \bigg [-{Q^2[(y-y_0)^2+(x-x_0)^2]\over 2}     \bigg ] 
   \label{30}
  \end{equation}
    \end{widetext}

Thus, when, upon moving of the sample in the $y$-direction, dependence of the HSS  $\delta u(x,y)$ reveals behavior of the type (\ref{30}), \cite{note6}, it indicates the presence of a localized inhomogeneity. The volume to susceptibility product, for this inhomogeneity, can be estimated using Eq. (\ref{30}):

\begin{equation}
\begin{split}
A_0={2h kQ^2 \alpha_0 S_0}\sqrt{ W_t\over W}=\sqrt{ W_t\over W}{16\pi \over \lambda \rho_c^2}V\alpha_0\Rightarrow \\ 
\alpha_0V=A_0\sqrt{ W\over W_t}{\lambda \rho_c^2\over 16\pi}
\end{split}
\end{equation}

  Here, $V\equiv S_0h$ is the scatterer volume. 
   \begin{figure}
               \includegraphics[width=.8\columnwidth,clip]{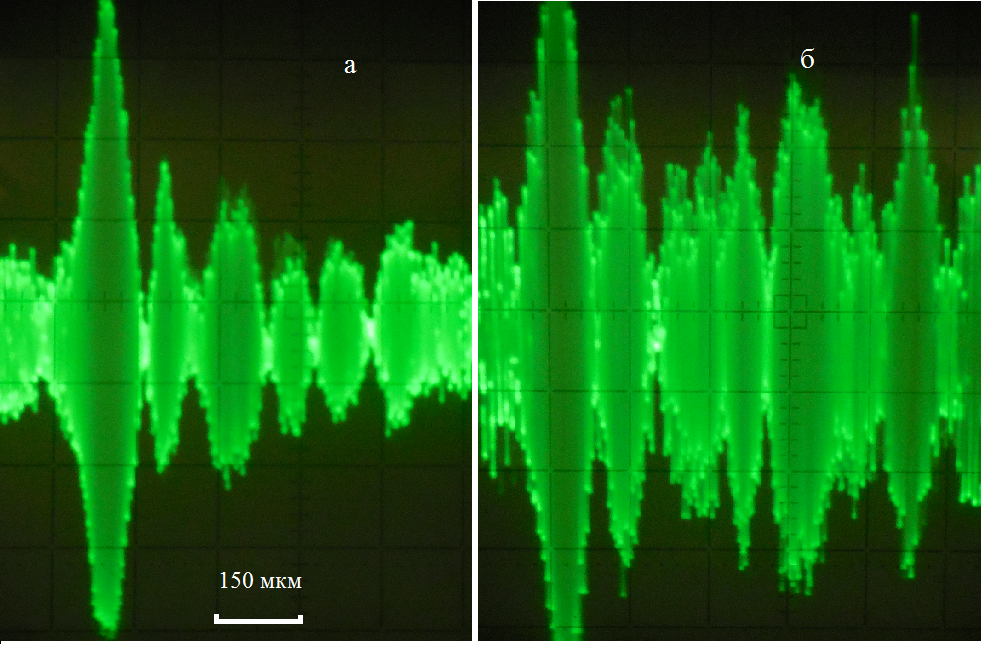}
                \caption{
                Oscillograms of the HSS $\delta u(t)$ obtained from a thin glass plate mounted on the vibrator oscillating in  the plane of the main and tilted beams. Two types of scatterers are presented: localized (a) and extended (b).  
                }
                \label{fig2} 
               \end{figure} 
  As the second example, let us consider a thin plate extended in the $xy$-plane uniformly translated with respect to the beam overlap region in the $y$-direction with the velocity $v$.  Inhomogeneity of the plate is simulated by a set of identical point scatterers randomly arranged in the $xy$-plane with the mean surface density $\sigma $. In this case, the HSS will depend on time $\delta u\rightarrow \delta u(t)$, and for its calculation we have to set $y = vt$ in Eq. (\ref{29}). With no loss of generality, we can assume that $x = 0$. By analogy with the previous example, we can write, for the susceptibility of the plate, the expression:
    
  \begin{equation}
  \alpha_{xx}(x,y)=S_0\alpha_0\sum_{i=1}^N\delta (x-x_i)\delta (y-y_i)
  \end{equation} 
  
  Here, $x_i$ and  $y_i$ are independent random quantities uniformly distributed over the intervals $[-X/2,X/2]$ and  $[-Y/2,Y/2]$, respectively ($X,Y\gg 2/Q=\rho_c$ are dimensions of the plate in the $x$ and $y$ directions), $N$ is the total number of scatterers with $N/XY=\sigma$.  Now, the Eq. (\ref{29}) acquires the form
  
  \begin{widetext}
   
  \begin{equation}
  \delta u(t)= -{2h kQ^2}\sqrt{ W_t\over W}\hskip1mm
  S_0\alpha_0\hskip1mm \sum_{i=1}^N  e^{-Q^2x_i^2/2}
      \sin \bigg [k(vt-y_i)\Theta \bigg ] \exp \bigg [-{Q^2(vt-y_i)^2\over 2}     \bigg ]
   \label{106}
  \end{equation}
\end{widetext}

One can see that $\delta u(t)$, in this case, represents a random process. Standard calculation of the correlation function of this process $\langle \delta u(t)\delta u(0)\rangle $ yields the following result:

\begin{equation}
      \langle \delta u(t)\delta u(0)\rangle=
      8[h k S_0\alpha_0]^2{ W_t\over W}\hskip1mm
        {\pi \sigma \over \rho_c^2}
      \hskip1mm \exp \bigg (-{v^2t^2\over \rho_c^2}\bigg )\hskip1mm \cos [k\Theta vt]
     \label{34}
     \end{equation}

 Equation (\ref{34}) shows that the random process $\delta u(t)$, in the considered case, is spectrally localized in the vicinity of the frequency $\Omega_0=k\Theta v$ and represents random oscillations at this frequency with characteristic amplitude $A_0\sim \sqrt{\langle \delta u(0)\delta u(0)\rangle}$.  This amplitude can be easily estimated in a real experiment and used to express parameters of the considered model of random scatterer:

 \begin{equation}
 \begin{split}
       A_0\sim \sqrt{\langle \delta u(0)\delta u(0)\rangle}=
       {4h k S_0\alpha_0\over \rho_c}\sqrt{2 W_t \pi \sigma\over W}\hskip1mm\Rightarrow
       \\ S_0\alpha_0\sqrt{\sigma}\sim {A_0\over 4}\sqrt{W\over 2\pi W_t}\hskip1mm {\rho_c\over hk}
       \end{split}
       \end{equation}

 Figure \ref{2} shows oscillograms of the HSS $\delta u(t)$ obtained from a thin glass plate using the above method. They, indeed, show oscillations at the frequency $\Omega_0=k\Theta v$ and often, at least qualitatively, can be referred to one of the two types of scatterers considered above: localized (Fig. \ref{fig2}a) and extended (Fig. \ref{fig2}b).  In the next section, we present results of a simple experiment on mapping of a quasi-plane sample.

\section{Experimental illustration}

   \begin{figure}
             \includegraphics[width=.8\columnwidth,clip]{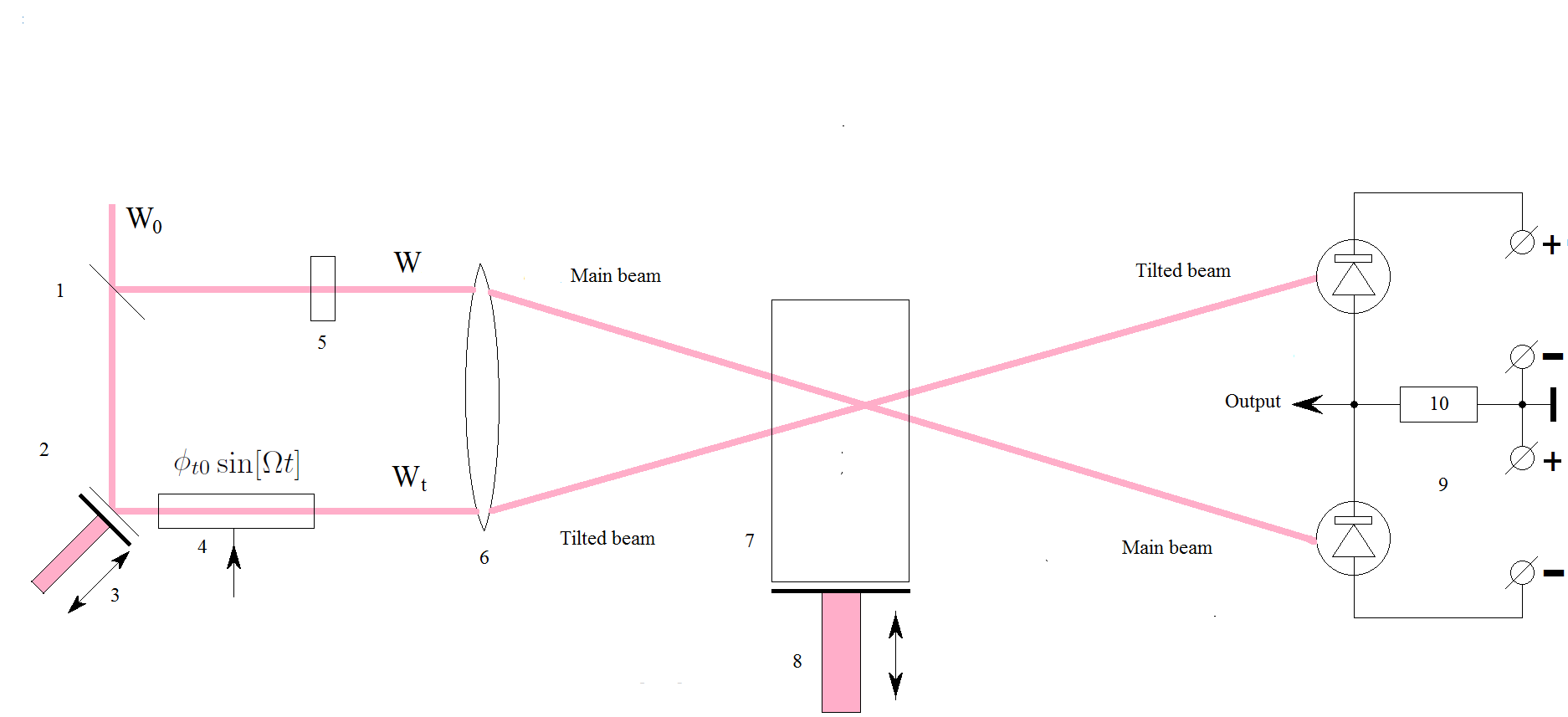}
              \caption{Heterodyne detection of scattering with subtraction of signals of the two beams. 1 - beamsplitter, 2 - mirror, 3 - vibrator 4 - Pockels cell, 5 - attenuator, 6 - lense 7 - sample, 8 - vibrator, 9 -  two-channel differential detector}
              \label{fig3} 
             \end{figure}

   Schematic of the setup for detection of the HSS and observation of properties of this signal is shown in Fig. \ref{fig3}. The main and tilted beams intersecting inside the sample 7 are split from the initial laser beam ($W_0\sim 2 - 3$ mW, $\lambda =650$ nm) using beamsplitter 1, mirror 2, and lens 6 ($f=100$ mm).  In the treatment presented above, the main and tilted beams were not equivalent – the tilted beam produced the scattering, while the main one served as a local oscillator for detection of the scattered field. The two-channel differential detector 9 allowed us to get use of symmetry of the main and tilted beams, due to which the scattered field created by the main beam produced HSS in the channel of the tilted beam (upper photodetector in Fig. \ref{fig3}).  So, the roles of the tilted and main beams, in this case, are interchanged. For this reason, when calculating the HSS in the channel of the tilted beam, one has to interchange, in Eq. (\ref{18}), the amplitudes of the tilted and main beams. This will lead to complex conjugation of the scalar product in this equation, and, as a result, the heterodyned scattering signals detected in the channels of the main and tilted beams appear to be equal in magnitude and opposite in sign. One can use both signals with the help of the differential photodetector shown in Fig. \ref{fig3}, which makes it possible not only to increase twice the observed HSS, but also to get rid of excess noise of the laser beam $W_0$.  The phase modulation of the tilted beam mentioned in the previous section, that allows one to distinguish HSS on the background of spurious signals, is performed either using mirror 2 on vibrator 3, or by means of  Pockels cell 4.

   \begin{figure}
                  \includegraphics[width=.8\columnwidth,clip]{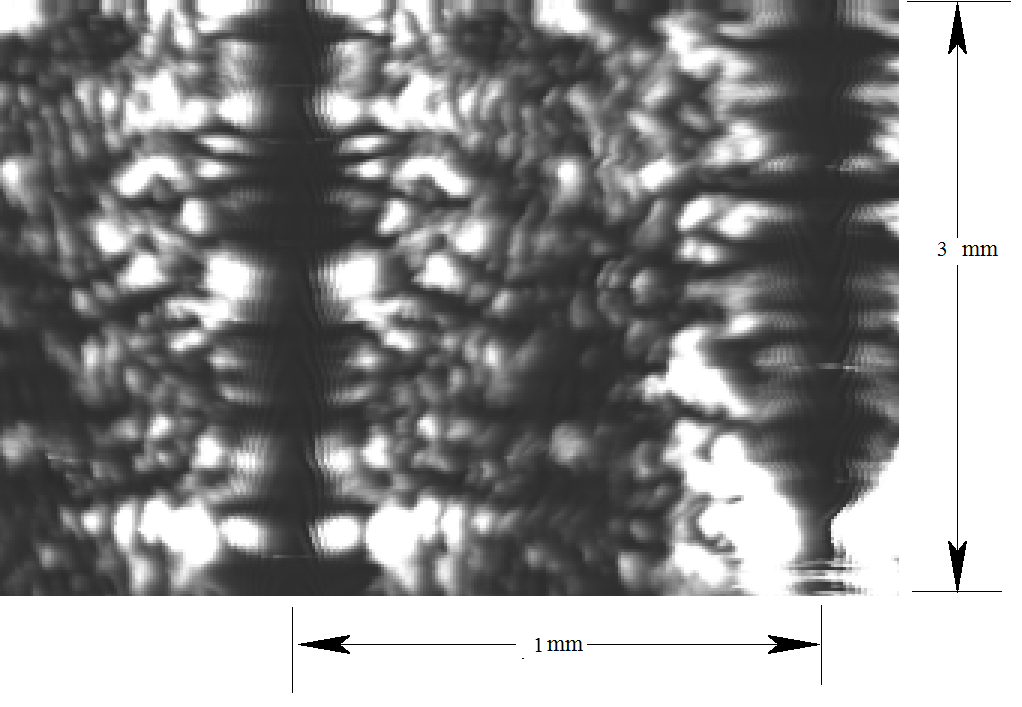}
                   \caption{Mapping of a plane scatterer.}
                   \label{fig4} 
                  \end{figure}

  Vibrator 8 provided displacement of the sample  $\sim$ 1 mm and served for mapping described in the previous section.  The result of such a mapping is presented in Fig. \ref{fig4} and was obtained in the following way.  The sample (a thin plate of silicate glass positioned in the region of the beam overlap) was fixed on vibrator 8 oscillating in the plane of the beams (see Fig. \ref{fig3})) with the amplitude 0.5 mm and frequency 60 Hz. Under these conditions, at the output of the differential detector 9 (on resistor 10), there has been detected the HSS that represented oscillations at the frequency $k\Theta v$  (see Eq. (\ref{34}) and Fig.\ref{fig2} a,b) with varying amplitude. These variations reflected inhomogeneity of the plate along the direction of the sample displacement.  Dependence of amplitude of these vibrations on displacement of vibrator 8 was recorded into the computer memory as a row of a 2D array (we recorded 300 counts). Then, we performed a small displacement in the direction orthogonal to the plane of the beams, and the next row was recorded, and so on.  The 2D array of 200 rows was displayed on the monitor of the computer in the form of a relief of brightness (Fig. \ref{fig4}).  In accordance with the model picture presented in the previous section, the isolated bright regions with dimensions $\sim 2\rho_c=60 \mu$m  can be interpreted using the first of the above examples (a small isolated scatterrer), while the extended regions with relatively small amplitude of the HSS – using the second example (a cluster of randomly arranged scatterers).  

The pattern presented in Fig. \ref{fig4}  serves only for illustration of the results of the above treatment, and we will not analyze it in more detail.  Note only that the dimensionless HSS $|\delta u|$, in these experiments, was of the order of $10^{-3}$ and could be easily detected experimentally. The double amplitude of the vibrator oscillations was 1 mm (in Fig. \ref{fig4}, it is shown by a horizontal two-headed arrow).  Displacement in the orthogonal direction was aroud 3 mm. One can easily see, in Fig. \ref{fig4}, the turning points  of  the vibrator and subsequent mirror replica of the pattern. One can also notice distortion of scale in the vicinity of these points, because our vibrator performed sinusoidal, rather than saw-wise, vibration.
 
Note that on the setup shown in Fig. \ref{fig3}, the HSS could be observed practically from any transparent scatterer placed into the region of the beam overlap.  In particular, we easily detected the HSS from a cuvette filled with pure water. In this case, we modulated the phase of the tilted beam using the Pockels cell, and the detected temporal fluctuations of HSS reflected the convective motion of suspended particles and random temperature-related variations of refractive index of the water. The experiments on volume tomography, which require precise 3D positioners, lie outside the scope of this paper.  Still, the above conclusion that the HSS is formed in the 3D region of overlap of the main and tilted beams was confirmed by the observation that when the cuvette was shifted in the direction of the $z$-axis, we could pass from the HSS fluctuating in time (when the overlap region was inside the water) to the stationary HSS (when it shifted to the glass wall of the cuvette).

   \section{ Conclusions}

     In this paper, we have calculated the signal detected with a flat photodetector irradiated by a strong optical beam of constant intensity (local oscillator) and a relatively weak signal field.  We show that when the signal field is a result of linear scattering of the additional tilted beam by a transparent inhomogeneous object (illuminated also by the light of local oscillator), the detected signal is determined only by the part of the scatterer that lies in the region of beam overlap.  We also show that observation of such signals allows one to solve problems of tomography of optically inhomogeneous transparent objects. Experimental illustration of tomography of a thin glass plate is presented.

   \section*{Acknowledgments}
  
  The author acknowledges financial support  from  the Russian Science Foundation (grant No. 17-12-01124).

\newpage



\begin{thebibliography}{99}
\bibitem{note1} This follows from transversality of electromagnetic wave. Changes of polarization at the edges of the beam are neglected.
\bibitem{note2}  Generalization for the case of tensor susceptibility will be presented below.
\bibitem{note5} Of course, these beams are supposed to be coherent.
\bibitem{note6} Shaped as an isolated peak of the amplitude $A_0$ and width $\sim 2/Q=\rho_c$, filled with oscillations with the spatial period $\sim 2\pi/k\Theta=\lambda/\Theta $ (which is often observed upon scanning of this glass plates, see Fig. 2a).
\end{thebibliography}
\end{document}